\documentclass[12pt,preprint]{aastex}

\slugcomment{}

\shorttitle{Milli-magnitude Precision Photometry}
\shortauthors{L\'opez-Morales}

\begin{document}

\title{Milli-magnitude Precision Photometry of Southern Bright Stars with a 1-m Telescope and a Standard CCD}

\author{Mercedes L\'opez-Morales\altaffilmark{1}}
\affil{Carnegie Institution of Washington, Department of Terrestrial Magnetism,
       5241 Broad Branch Rd. NW, Washington D.C., 20015, USA}
\email{mercedes@dtm.ciw.edu}

\altaffiltext{1}{Carnegie Fellow.}

\begin{abstract}
This paper summarizes a three night observing campaign aimed at achieving milli-magnitude precision photometry of bright stars (V $<$ 9.0) with the 1-meter Swope telescope at Las Campanas Observatory. The test targets were the main sequence stars HD205739 and HD135446. The results show that, by placing a concentric diaphragm in front of the aperture of the telescope, it is possible to avoid saturation and to achieve a photometric precision of 0.0008--0.0010 mag per data point with a cadence of less than 4 minutes. It is also possible to reach an overall precision of less that 0.0015 mags for time series of 6 hours or more. The photometric precision of this setup is only limited by scintillation. Scintillation could be reduced, and therefore the photometric precision could be further improved, by using a neutral density filter instead of the aperture stop. Given that the expected median depth of extrasolar planet transits of about 0.01 mags, and their typical duration of several hours, the results of this paper show that 1-m telescopes equipped with standard CCDs can be used to detect planet transits as shallow as 0.002 mags around bright stars.

\end{abstract}

\keywords{techniques: photometric --- stars: planetary systems.}

\section{Introduction}\label{sec:intro}

High precision photometry of bright stars is becoming a subject of increasing interest on extrasolar planet studies. Precision photometry complements the Doppler observations by establishing whether the observed radial velocity variations are caused by the reflex motion induced by a planetary companion or by stellar magnetic activity (e.g., Henry et al. 2000a, Queloz et al. 2001, and Paulson et al., 2004). Precision photometry can also determine the presence or absence of transits in stars with known planets. The presence of a transit provides additional information about the planet by allowing the determination of its radius and its absolute mass (e.g., Charbonneau et al. 2000, Henry et al., 2000b, Sato et al. 2005, and Charbonneau et al. 2005), and also more precise orbital parameters of the system.

Photometric precision of a few milli-magnitudes is routinely achieved using ground-based telescopes and standard CCDs for stars fainter than V = 9.0--10.0. Gilliland et al.(1991) showed that, using a 1-m telescope and a standard CCD, it is possible to achieve a precision of 0.0008 mags for 13th magnitude stars, in time series of about 9 hours with 2-minute candence. Everett \& Howell (2001) have more recently demonstrated that it is possible to reach precisions of 0.00019 mags in a 4.5-hour time series for V = 14.0 stars with a 0.9-m telescope and a mosaic CCD. However, for stars V = 9.0 mag or brighter milli-mag precision photometry basically remains the realm of photomultiplier tubes\footnote{The most successful ground-based experiment using photomultipliers is the automatic photometric telescopes (APTs) at Fairborn Observatory (Henry 1995a, Henry 1995b, Stammeir et al. 1997, Henry 1999, and Eaton, Henry \& Fekel 2003). The APTs routinely achieve photometric precision of 0.001 mag for single observations of stars brighter than V = 8.5. They are also the only instruments that have maintained that precision for more than ten years, what makes them a superb tool.}. A typical CCD inmmediatly saturates when we place it in a 1-m class telescope and try to image a bright star. Saturation can be avoided by using a smaller telescope, but in that case atmospheric scintillation limits the photometric precision to 3--5 millimags.

Efforts to reach milli-mag precision photometry of bright stars with CCDs on ground based 1-m class telescopes are now beginning to produce results comparable to those of photomultipliers tubes. For example, Bouchy et al. (2005) recently published the detection of an extrasolar planet transit around HD189733 with the 1.2 m telescope at the Haute-Provence Observatory in France, using a 1024 x 1024 SITe back-illuminated CCD. The depth of this transit is 0.03 mags and their photometric precision per data point varies between 0.002 and 0.003 mags. Similarly, Charbonneau et al. (2006) have confirmed the 0.003 mag deep transit of HD149026b (Sato et al. 2005) using a 2-chip 2048 x 4608 CCD installed at the 48-inch (1.2m) telescope of the F. L. Whipple Observatory at Mount Hopkins, Arizona. Most efforts have been focused on the Northern Hemisphere, where small telescopes are more at hand. However, at least half of the sample of stars currently searched for planets lie at southern latitudes. In this paper I present the results of tests performed at the 40-inch (1m) telescope in Las Campanas Observatory, Chile. The tests were intended to achieve milli-mag precision photometry of southern bright currently being searched for planets by the radial velocity surveys. The selected targets were HD205739 (V = 8.5) and HD135446 (V = 8.2). Section 2 describes the instrumental setup and the observations. Section 3 summarizes the data analysis and presents the resulting light curves. Section 4 compares our results to the three currently known exoplanet transits around bright stars. Finally the summary of this work and future plans are given in section 5.

\section{Instrument Setup and Observations}\label{sec:setobs}

\subsection{Instrument Setup}\label{sec:setup}

The data was collected at the Henrietta Swope telescope, located in Las Campanas Observatory in Chile. The Swope is a 40-inch (1m) reflector with an f/7 Ritchey-Chr\'etien optical design. It is currently equipped with a 2048 x 3150, 15 $\mu$m pixel, SITe CCD that provides an unvignetted FoV of 14.8 x 22.8 arcminutes. The dynamic range of the CCD is 32727 ADU and it converts electrons into ADUs at a fixed gain of 2.5 $e^{-}/ADU$. The readout time of the CCD on its 1 x 1 unbinned configuration is 128 sec. In addition, there is a minimum exposure time limit of 5 sec to ensure that the shutter has moved completely out of the way.

With that setup stars brighter than V= 8.5 will unaviodably saturate in the 5 sec minimum exposure time. Saturation occurs even if we try to distribute the photons over a larger number of pixels by defocusing the telescope.

One way to prevent saturation is to use a neutral density filter to block away a fraction of the photons arriving from the stars. Unfortunately no neutral density filter is currently available at Swope. Another way to reduce the number of photons impacting on the CCD is to place an aperture stop to reduce the photon collecting area of the telescope.

\begin{figure}[t]
\epsscale{1.0}
\plotone{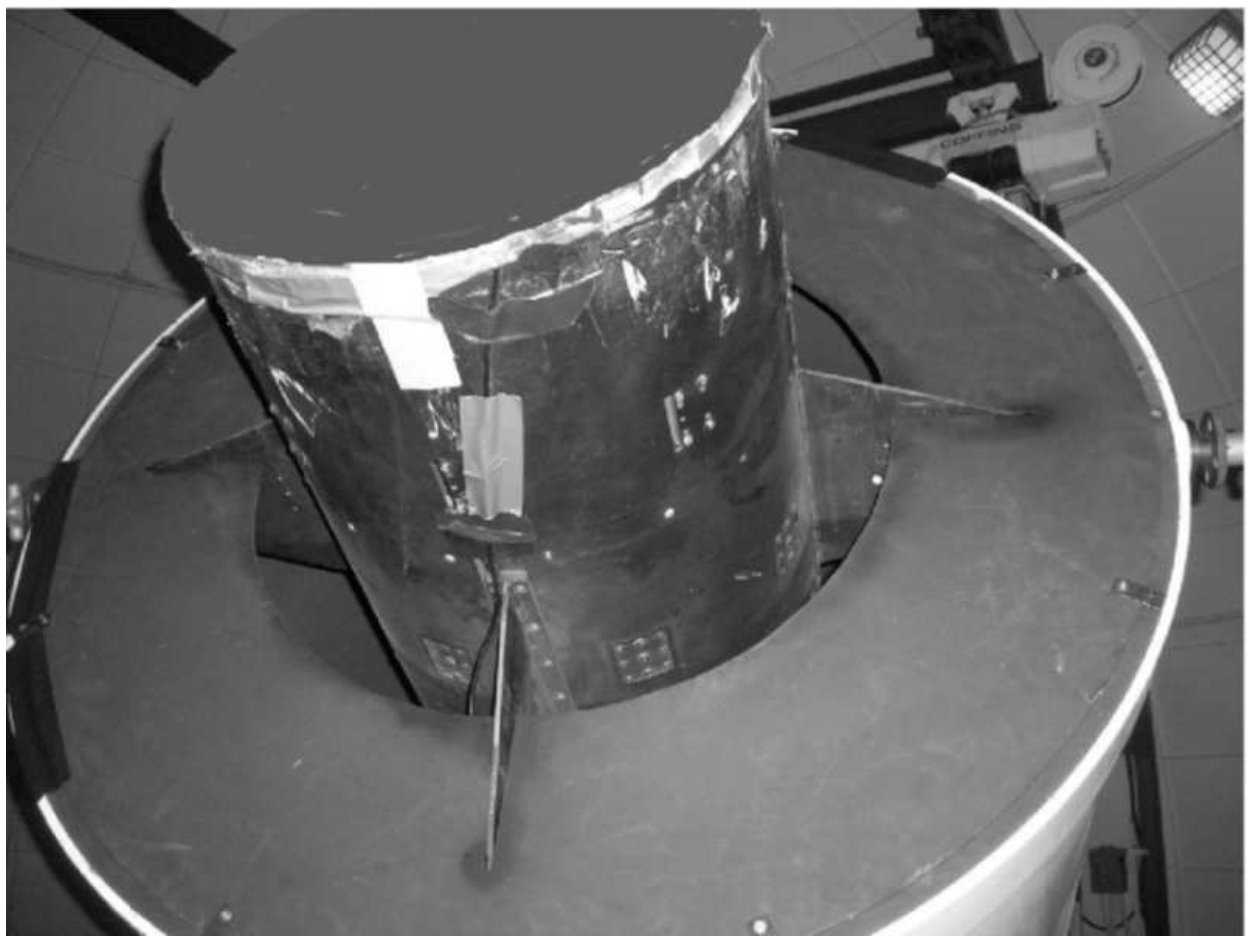}
\caption{Aperture stop to reduce the effective aperture of the Swope from 0.589 $m^2$ to 0.146 $m^2$. The stop is a 1 meter in diameter piece of plywood with a central aperture of diameter 0.67 m. The stop reduces the brightness of the stars by two magnitudes, so it is possible to image stars as bright as V=6.5 without saturating.}
\label{fig:tel}
\end{figure}

With the help of Las Campanas Observatory personnel, I built a stop to reduce the effective aperture of the telescope from 0.589 $m^2$ to 0.146 $m^2$. The stop, a 1 meter in diameter piece of plywood painted in black and with a central aperture of diameter 0.67 m, was placed in front of the aperture of the telescope as shown in figure \ref{fig:tel}. Objects appear 2.0 magnitudes fainter with that setup, making it possible to image stars as bright as V = 6.5 without saturating.

To reduce in some amount the duty cycle of the observations, the readout of the CCD was limited to an area of 2048 x 2048 pixels (14.8 x 14.8 arcmin). That area is still large enough to simultaneously image the target star and at least one nearby comparison star of similar magnitude, but reduces the readout time to 90 sec, instead of the 128 sec that takes to read out the full CCD.

\subsection{Observations}\label{sec:obs}

The test targets for this work were the two southern stars HD205739 (V = 8.57; B--V = 0.50; Sp.~Type: F7V) and HD135446 (V = 8.20; B--V = 0.57, Sp.~Type: G1.5V). Both stars have no previous record of any kind of intrinsic variability and are being searched for planets by the radial velocity surveys. HD205739 was observed during three nights, on UT 2005 Aug 3--4, Aug 8--9, and Aug 9--10. HD135446 was observed on the nights of UT 2005 Aug 8--9 and Aug 9--10. A log of the observations is presented in Table \ref{tab:ObsLog}. HD135446 was monitored over the first three hours of each night. HD205739 was monitored over the second half of the nights (about 6 hours per night). The observations were all collected under photometric conditions, spanning over air masses between 1.01 and 1.63.

The star used as comparison to HD205739 was HD205860, a star slightly brigther than HD205739 and of the same spectral type (V = 8.27; B--V = 0.50, Sp.~Type: F7V). This last point is very fortunate, since it eliminates from the differential photometry the second order extinction effects introduced by differences in color between the comparison and the target stars. The separation between the two stars is $\Delta\alpha$ = $46^s$.32 and $\Delta\delta$ = - 10'.31. In the case of HD135446 the comparison was HD135342 (V = 9.27; B--V = 0.51, Sp.~Type: F6V). The separation between target and comparison is now $\Delta\alpha$ = $-25^s$.04, and $\Delta\delta$ = 1'.87. Both stars have also similar colors in this case, but the comparison is 1.1 magnitudes fainter than the target.

The telescope was defocused so that the count level per pixel in both the target and comparison stars remained within the linearity limit of the CCD ($<$ 25,000 ADU). The FWHM of the stars during the observations range between 15 and 19.5 pixels (6.5--8.5 arcsec) in the images of HD205739 and between 19 and 23.5 pixels (8.2--10.2 arcsec) in the images of HD135446. While keeping the count levels below 25,000 ADU, the exposure times also had to be long enough to 1) ensure that enough photons were collected in both the targets and the comparison stars to keep the Poisson noise level below one part in a thousand, and 2) to reduce the errors introduced by scintillation. Typical exposure times were 20--22 seconds for HD205739 and 30--32 seconds for HD135446, respectively.

Pixel to pixel sensitivity differences is one of the main precision limiting factors in CCD photometry. To minimize this contribution to the noise, the stars were placed over the same pixels each night, and then monitored with careful telescope guiding to ensure that the center of the stars was always located within the same $\pm$ 1--2 pixels. All the images were taken using a V-band Johnson standard filter.

Finally, several hundred bias and flat frames were collected each night to ensure an overall count level of over $10^6$ photo-electrons per pixel in the final calibration frames. In this manner one can avoid introducing additional random noise in the calibrated images beyond the 1/1000 intended photometric precision.

\section{Analysis and Results}\label{sec:anal}

All the images were bias substracted and flat-fielded using the same combined bias and combined flat frames. Those frames were generated by combining all the biases and flats collected in different nights, so that each of the final two calibration frames had an overall count level of at least $10^6$ photons. That level of counts ensures that the random noise introduced during the calibration of the images, assumed to be Poisson noise, does not exceed 0.001 mags.

The next step of the reduction processs consisted on performing aperture photometry of each target and their comparison star on the calibrated images. In order to optimize the results of the aperture photometry, I performed the photometry over series of apertures with radii varying between 12 and 35 pixels. The area used to compute the sky background around each star was always the same (an annulus of 15 pixels at a radius of 60 pixels centered on the stars). 

\begin{figure}[t]
\epsscale{1.0}
\plotone{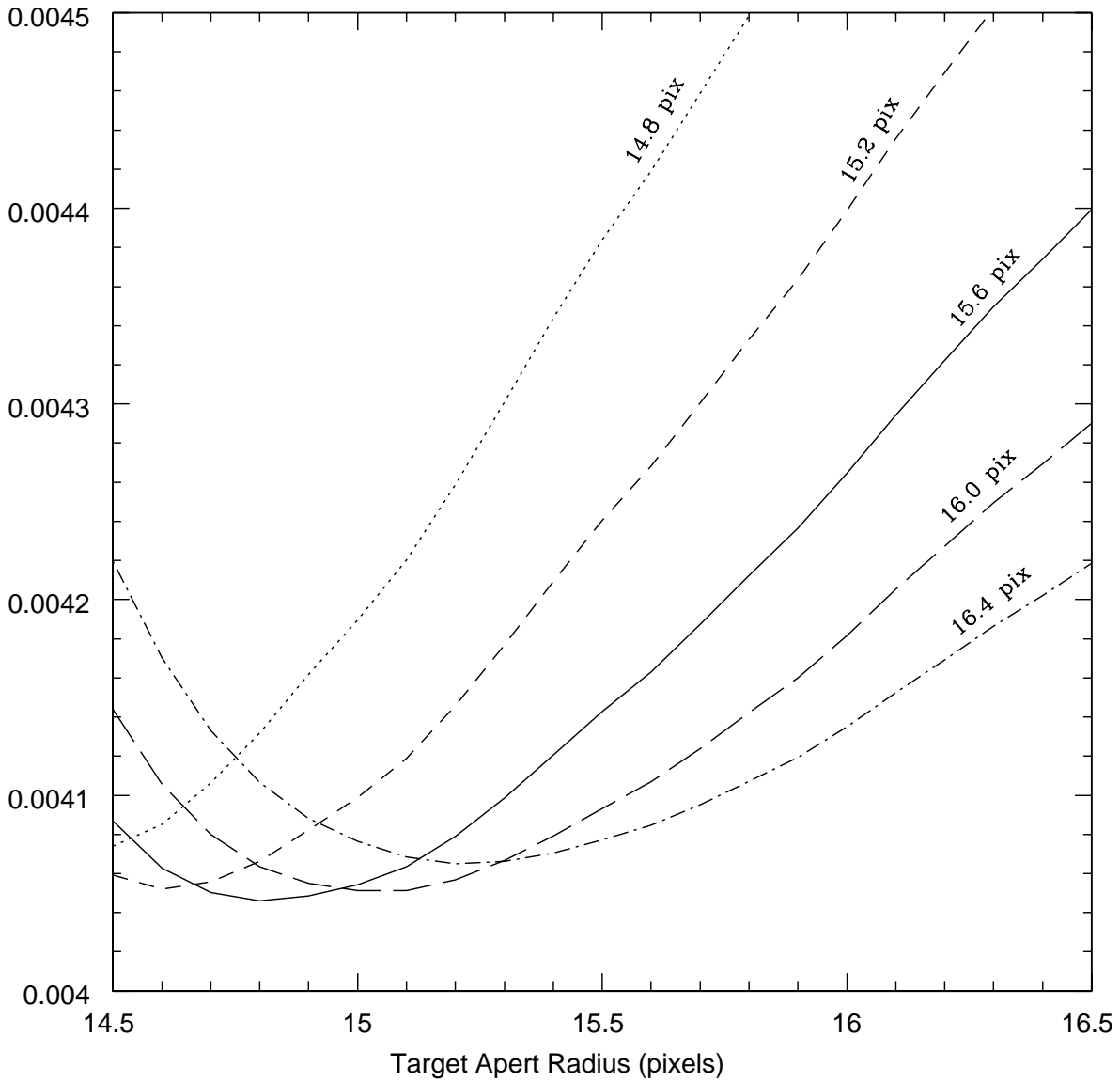}
\caption{Example of the aperture vs. $\sigma_{ft/fc}$ plots to determine the best combination of photometric apertures for the target and comparison stars, as described in section \ref{sec:anal}. The plot shows the values of $\sigma_{ft/fc}$ (y-axis) for different apertures around the target HD205739 (x-axis). Each line corresponds to a different aperture for the comparison star. The best photometry in this case is obtained when using an aperture of 14.8 pixels for the target and 15.6 pixels for the comparison star. The values of $\sigma_{ft/fc}$ in this plot include outlayers (points that deviate by more than 5$\sigma$ from the mean of the light curve). Those outlayers were removed from the final light curves.}
\label{fig:dmag}
\end{figure}

Differential photometry light curves were generated iteratively for all the possible combinations of apertures of the target and the comparison star. The best light curve was then selected as the one resulting from the combination of apertures that gave the smallest value of the standard deviation $\sigma_{ft/fc}$, where $f_{t}$ and $f_{c}$ are respectively the instrumental flux counts of the target and the comparison star. First, a prelimimary minimum of $\sigma_{ft/fc}$ is sought by assigning aperture increments of 1 pixel per iteration. Once an approximate value of the minimum of $\sigma_{f_{t}/f_{c}}$ is found, one runs a second pass around that minimum to optimize the solution. In that second pass the aperture increments are 0.1 pixels instead of the 1.0 pixels in the first pass. The optimum apertures found were different for each target and each night. An example of the aperture versus $\sigma_{ft/fc}$ plots generated to determine the best combination of apertures is illustrated in figure \ref{fig:dmag}, where I show the results of the aperture photometry of HD205739 for the data collected on Aug 3--4. The optimum apertures in that case were 14.8 pixels for the target and 15.6 pixels for the comparison star. The optimum apertures for each target in all the other nights are listed in Table \ref{tab:apert}.

\begin{figure}[t]
\epsscale{1.0}
\plotone{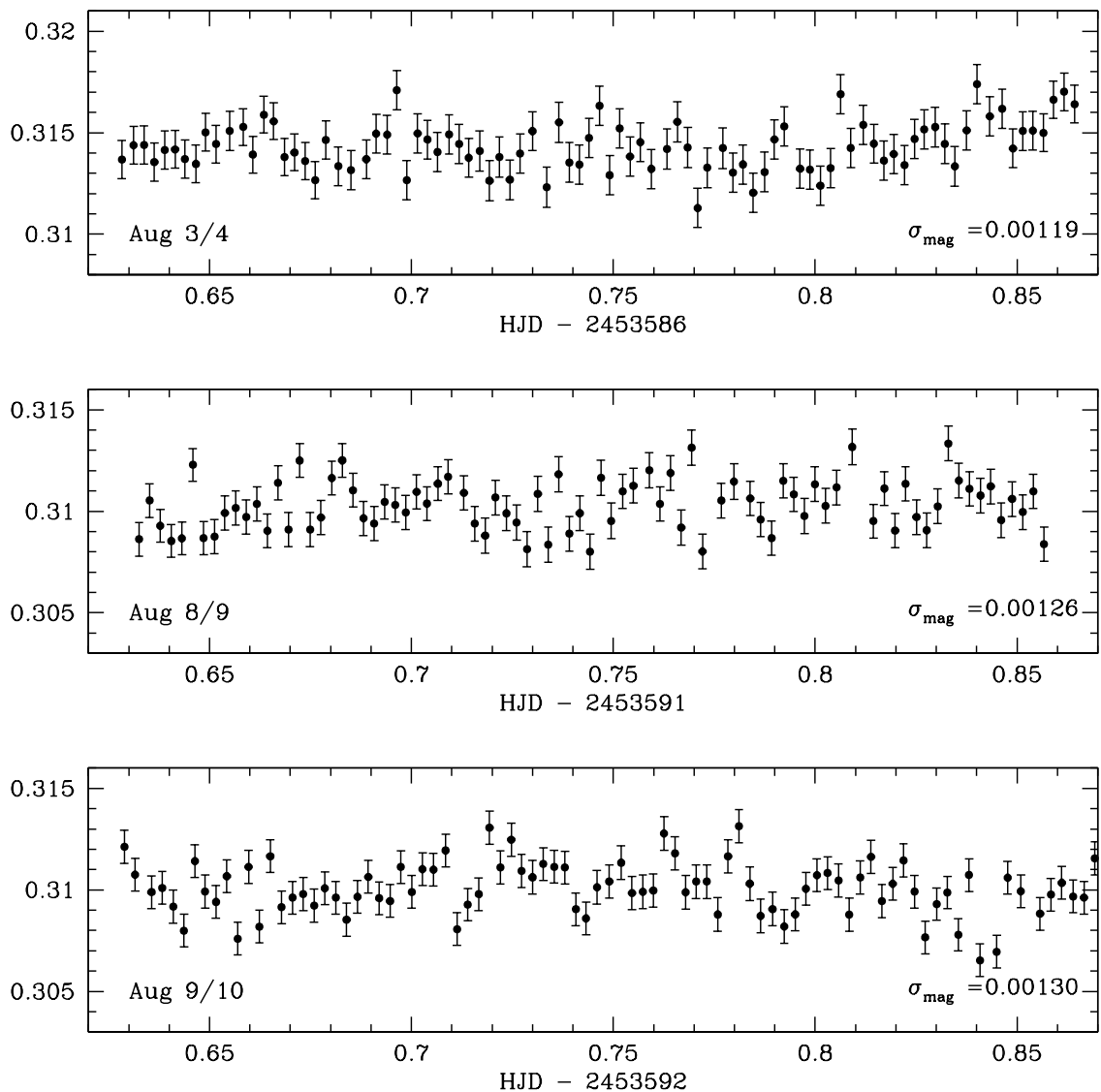}
\caption{16.95 hours of photometric coverage of HD205739 over three nights. The top, middle and bottom light curves corresopnd, respectively, to the data collected on UT 2005 Aug 3--4, Aug 8--9, and Aug 9--10. The average rms of the individual points range between 0.0008 and 0.0010 mags. The standard deviations of the nightly light curves are, from top to bottom, 0.00119, 0.00126, and 0.00130 magnitudes. The data have been binned into 2-points bins resulting on a time of cadence of 3.5 minutes.}
\label{fig:hd205739}
\end{figure}

\begin{figure}[t]
\epsscale{1.0}
\plotone{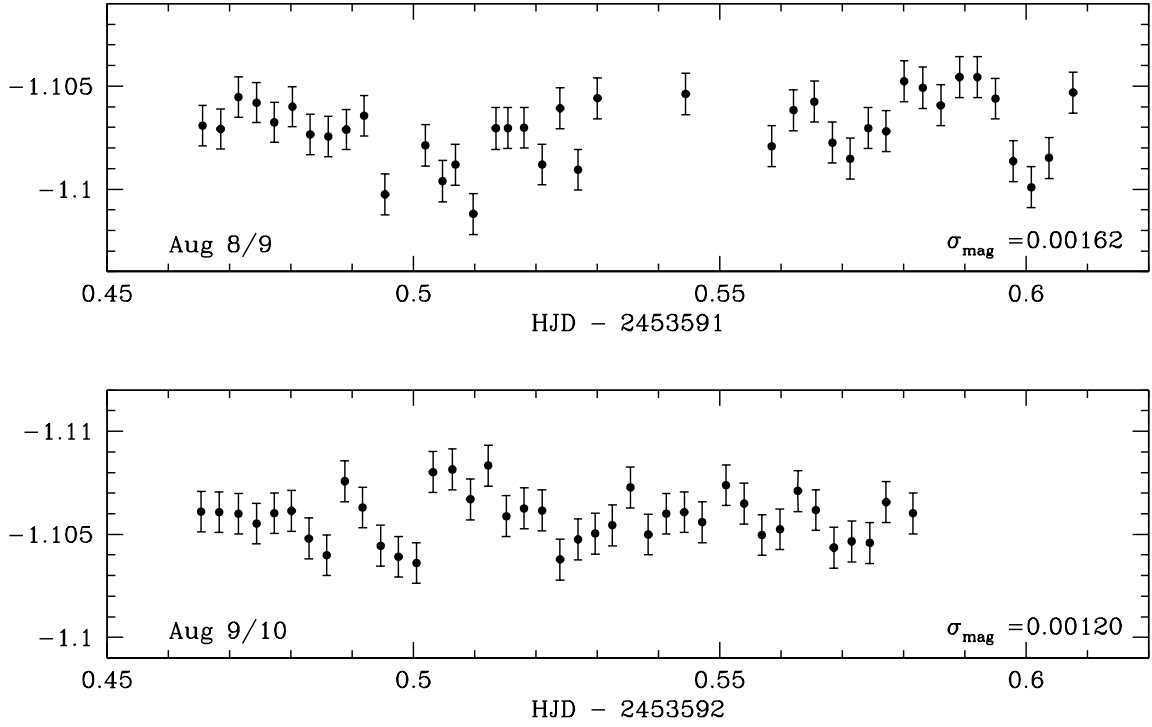}
\caption{6.2 hour photometric coverage of HD135446 over two nights, Aug 8--9 (top) and Aug 9--10 (bottom). The average rms of the individual points range between 0.0008 and 0.0010 mags. The standard deviations of the nightly light curves are respectively 0.00162 (top) and 0.00120 magnitudes (bottom). The data have been binned into 2-points bins resulting on a time of cadence of 4.0 minutes.}
\label{fig:hd135446}
\end{figure}

The final light curves are presented in figures \ref{fig:hd205739} and \ref{fig:hd135446}. Figure \ref{fig:hd205739} shows 16.95 hours of photometric coverage of HD205739 over the three nights that this star was observed (Aug 3--4, Aug 8--9, and Aug 9--10). Figure \ref{fig:hd135446} shows the 6.2 hours of coverage of HD135446 during Aug 8--9 and Aug 9--10. The data in both figures have been averaged into 2-point bins, resulting on a time of cadence of 3.5 minutes for HD205739 and 4.0 minutes in the case of HD135446.

The average rms of the individual points range between 0.0008 and 0.0010 mags, depending on the night. The average standard deviation of the nightly light curves range between 0.00119 and 0.00130 mags for HD205739 and 0.00120 and 0.00162 for HD135446, indicating that both targets and their comparison stars were photometrically stable to at least those levels of precision during the nights that they were observed.

The similarity in brightness and color of the targets to their comparison stars minimizes the appearance of differential extinction effects. These extinction effects are also reduced by limiting the photometric follow-up to airmasses smaller than 1.65. None of the light curves present differential extinction trends. I believe that the slight 0.003 mag slope between times 2453586.80 and 2453586.87 HJD in the top light curve in figure \ref{fig:hd205739} is real, after discarding extinction or background brightness gradient effects in the frames. Unfortunately, there are no follow-up observations to further investigate the cause of that slope.

Scintillation is the limiting factor to the photometric precision of this setup. Equation 10 from Dravis, Lindergren \& Mezey (1998) yields a scintillation photometric variation of 0.0012 mags for the average of two 25 sec exposures at an intermedian airmass of 1.3\footnote{Assuming a telescope aperture diameter of 66 cm, an observatory height of 2100 m, and an atmospheric scale height of 8000 m.}. That value is in agreement with the standard deviation of the light curves in figures \ref{fig:hd205739} and \ref{fig:hd135446}.

\section{Application to Extrasolar Planet Transits}

The expected median transit depth of giant close-in planets is 0.01 mags and the duration of those transits is typically of several hours. By comparing these numbers to the results presented in the previos section, one can conclude that a typical extrasolar planet transit would be easily detected around bright stars with the telescope setup and the observational plus data reduction strategies used Swope. 

\begin{figure}[t]
\epsscale{1.0}
\plotone{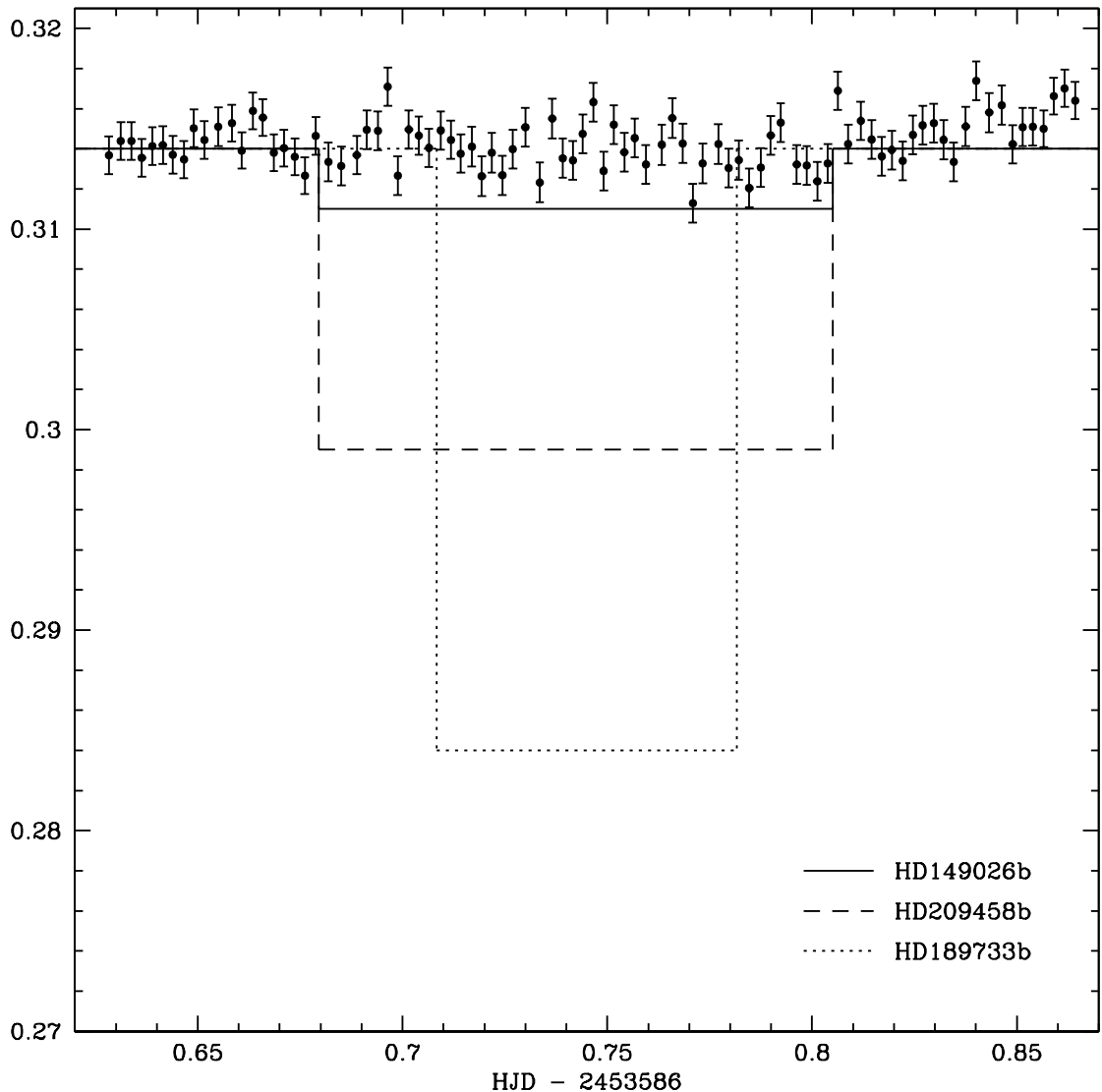}
\caption{Schematic representation of how the three known transits of extrasolar planets around bright stars (V $<$ 9.0) would look like when superimposed on one of our light curves. The continuous line represents the transit of HD149026b (Sato et al. 2005) and the dashed and dotted lines represent respectviely the transits of HD209458b (Charbonneau et al. 2000) and HD189733b (Bouchy et al. 2005). The values of $\Delta$t and $\Delta$mag for each transit are summarized in Table \ref{tab:deltas}.}
\label{fig:transits}
\end{figure}

To reinforce this statement, I provide in figure \ref{fig:transits} an schematic representation of how the three known transits of extrasolar planets around brigth stars (V $<$~ 9) would appear when supperimposed on one of the light curves obtained in this work. The continuous line simulates the transit of HD149026b (Sato et al. 2005), the dashed line simulates the transit of HD209458b (Charbonneau et al. 2000), and the dotted line simulates the transit of HD189733b (Bouchy et al. 2005). The transits have been represented by boxcar functions, instead of using a closer model to their real shape, to emphasize their duration $\Delta$t and their depth $\Delta$mag as compared to the observed light curves, represented by the dots. The values of $\Delta$t and $\Delta$mag have been estimated from the original papers and are summarized in Table \ref{tab:deltas}. 

The main conclusion from figure \ref{fig:transits} is that the three extrasolar planet transits would have beeing easily detected by the setup that we are using at Swope.

\section{Summary and Conclusions}

This paper shows the results of a three-day test campaign aimed at achieving milli-magnitude precision photometry of bright southern stars (V $<$ 9.0) with a 1-m telescope and a standard CCD. The targets for these tests were HD205739 and HD135446, two 8th magnitude stars with no previous record of photometric variability.

Stars that bright would inevitably saturate given the collecting area of the telescope and the minimum exposure time of 5 sec imposed by the displacement rate of the CCD shutter. In the absence of neutral density filters at Swope, saturation can be avoided by implementing an aperture stop to reduce the aperture of the telescope, dimming the stars by two magnitudes, and making it possible to image objects as bright as V = 6.5 without saturating.

The observations consisted of follow-up photometry of HD205739 over three nights, for a total of 16.95 hours. HD135446 was also monitored over two nights, for a total of 6.2 hours. Special care was taken to keep all sources of noise below the aimed photometric precision goal of 0.001 mags. The total number of counts in both the targets and the comparison stars was at least $10^6$ photo-electrons. The level of counts per pixel never exceeded the linearity limit of the CCD. The level of counts in the combined bias and master flat-field calibration frames was also at least $10^6$ photo-electrons. The stars were monitored only at airmasses under 1.65 to reduce differential extinction effects. The stars were also placed and kept over the same pixels each night to minimize errors caused by the pixel-to-pixel sensitivity variation of the CCD. Finally, we chose as comparison stars nearby objects with magnitudes and colors similar to the targets to minimize the effects of extinction. 

The aperture photometry was performed by iteratively generating differential photometry light curves for different combinations of apertures of the target and the comparison star. The best light curves were then selected as the ones resulting from the combination of apertures that gave the smallest standard deviations.

The resulting light curves show that it is possible to achieve precisions between 0.0008 and 0.0010 mags in individual points with a cadence of less than 4 minutes. The results also show that it is possible to keep the level of precision of the light curves below 0.0013-0.0015 for extended periods of time (at least 6 hours).

The precision of the light curves is limited by atmospheric scintillation, which with the diaphramed aperture of the telescope and the current exposure times (20-30 sec) introduces an average photometric variation of 0.0012 mags. Scintillation depends heavily on the diameter of the telescope. Its effect can be therefore reduced to 0.0008 mags if we replace the aperture stop by an inexpensive neutral density filter. The neutral density filter will have the same advantage of dimming the stars by 2 magnitudes, as the aperture stop, but will not have the inconvenience of the lose in telescope aperture that enhances the atmospheric scintillation.

These tests demonstrate that the Swope telescope can be used to obtain high precision photometry of southern bright stars, with very small alterations to its current setup. In particular Swope can be used to detect transits of close-in giant planets around bright planet-hosting stars. For example, Swope would had easily detected the transits of the planets HD189733b, with a depth of 0.030 mags, HD209458b, with a depth of 0.015 mags, and even the only 0.003-mag deep transit of HD149026b. There are currently no facilities in the Southern Hemisphere pursuing a targeted planet transit search among known planet-bearing stars. Swope is becoming therefore the first one of those facilities.

Future plans include the replacement of the aperture stop by neutral density filters and the extension of these tests to other wavelengths, mainly UBVRI Johnson filter passbands. Another technical goal is to try to improve the duty cycle of the observations. In addition, we are starting a targeted search for transits around the southern stars with known planets that have not been monitored yet.

\section{Acknowledgments} Thanks to Oscar Duhalde for his help building the telescope aperture stop. Many thanks also to Paul Butler, Alycia Weinberger, Kaspar von Braun and Alan Boss for their helpful commentaries on the composition of this document. The author acknowledges research and travel support from the Carnegie Institution of Washington through a Carnegie Fellowship.

\newpage

\begin{table}[t]
\begin{center}
\footnotesize
\caption{Log of observations of the test target stars HD205739 and HD135446}
\label{tab:ObsLog}
\vspace{0.3in}
\begin{tabular}{rccc}
\hline\hline
Date(UT)&Object&Time(HJD-2453500)&Airmass\\
\hline
2005 Aug 3/4&HD205739&86.62829--86.86426& 1.18-1.42\\
2005 Aug 8/9&HD135446&91.46557--91.60766& 1.01-1.63\\
             &HD205739&91.63257--91.85669& 1.11-1.51\\
2005 Aug 9/10&HD135446&92.46533--92.58154& 1.01-1.59\\
             &HD205739&92.62891--92.87488& 1.12-1.58\\
\hline\hline
\end{tabular}
\end{center} 
\end{table}

\begin{table}[t]
\begin{center}
\footnotesize
\caption{Optimum photometric apertures for each target in each night}
\label{tab:apert}
\vspace{0.3in}
\begin{tabular}{rcccc}
\hline\hline
Target&Date& Optimum Apert&Optimum Apert& $\sigma_{f_{t}/f_{c}}$\tablenotemark{\dagger}\\
      &(UT)& of Target (pix)&of Comparison (pix)&\\

\hline\hline
HD135446&Aug 8/9 2005&18.6&18.8&0.00506\\
        &Aug 9/10 2005&23.1&22.1&0.00501\\

\hline
HD205738&Aug 3/4 2005&14.8&15.6&0.00405\\
        &Aug 8/9 2005&19.2&18.9&0.00452\\
        &Aug 9/10 2005&18.8&18.4&0.00235\\

\hline\hline
\end{tabular}
\end{center}
\tablenotetext{\dagger}{The values of $\sigma_{ft/fc}$ in this table include outlayers (points that deviate by more than 5$\sigma$ from the mean value of the light curve). Those outlayers were removed from the final light curves.}
\end{table}

\begin{table}[t]
\begin{center}
\footnotesize
\caption{Duration ($\Delta$t) and depth ($\Delta$mag) of the transits of HD149026b, HD209458b and HD189733b}
\label{tab:deltas}
\vspace{0.3in}
\begin{tabular}{ccc}
\hline\hline
Planet&$\Delta$t (hours)& $\Delta$mag (mags) \\
\hline\hline
HD149026b& $\sim$ 3.0 & 0.003\\
HD209458b& $\sim$ 3.0 & 0.015\\
HD189733b& $\sim$ 1.76 & 0.030\\
\hline\hline
\end{tabular}
\end{center}
\end{table}

\newpage

\end{document}